\documentclass[aps,twocolumn,prl]{revtex4-1}%
\usepackage{amsfonts}
\usepackage{amsmath}
\usepackage{amssymb}
\usepackage{graphicx}
\usepackage{epsfig}
\usepackage{subfigure}
\usepackage{booktabs}
\usepackage{natbib}%
\providecommand{\U}[1]{\protect\rule{.1in}{.1in}}

\begin{document}
\title{Comment on \textquotedblleft Linear Scaling of the Exciton Binding Energy versus the Band Gap of Two-Dimensional Materials\textquotedblright\\ }
\author{Mingliang Zhang, Ling-Yi Huang, Xu Zhang and Gang Lu$^{\ast}$}
\affiliation{Department of Physics and Astronomy, California State University Northridge, Northridge CA91330}
\date{July 17, 2016}
\maketitle


In a recent Letter, Choi \textit{et al.} have performed first-principles
GW-Bethe-Salpeter equation (GW-BSE) calculations for a number of
two-dimensional (2D) semiconductors and discovered a linear scaling relation
between exciton binding energy $E_{b}$ and quasi-particle bandgap $E_{g}$
\cite{choi}. The authors further state that the linear scaling is expected to
be applicable to essentially all existing and future 2D materials. In this
Comment, we show that this linear scaling relation does not apply to all 2D
materials, and a deviation from the linear scaling is predicted for small
bandgap 2D materials.

We first note that the linear relation revealed in Fig. 4 of Choi's work
cannot extend to a vanishing $E_{g}$, because it would imply a negative
optical bandgap (the difference between $E_{g}$ and $E_{b}$). Instead, $E_{b}$
should vanish as $E_{g}$ approaches zero, deviating from the linear relation.
To support this claim, we have carried out the first-principles GW-BSE
calculations with essentially the same computational parameters as Choi
\textit{et al.} for a number of \textit{small bandgap} 2D semiconductors. The
computational details can be found in Supporting Material. Specifically, we
stretch the zero bandgap graphene with tensile strains to open small bandgaps,
and compress the 2D phosphorene to reduce its bandgap. The results are
summarized in Fig. 1 along with the original data points from Choi's paper.
First of all, we reveal that for small bandgaps ($E_{g} < 2$ eV), the linear
scaling relation is clearly violated, and as expected, $E_{b}$ drops to zero
much faster than what was predicted by the linear relation. Secondly, we
confirm that the linear scaling remains valid for 2D semiconductors whose
bandgap is greater than 2 eV. In fact, our data point of the largest $E_{g}$
coincides with that of Choi of the smallest $E_{g}$, which partially
corroborates the calculations of Choi \textit{et al.}

To shed light on the results, we have derived an analytic expression
correlating $E_{g}$ and $E_{b}$. Our analysis is based on the same hydrogenic
model as used in Choi's paper, where $E_{b}\varpropto\mu/\varepsilon^{2}$ is
assumed. $\mu=m_{e}m_{h}/(m_{e}+m_{h})$ is the reduced mass of the exciton,
and $m_{e}$ and $m_{h}$ is the effective mass of the electron and the hole,
respectively; $\varepsilon$ is the static dielectric constant of the 2D
semiconductor \cite{choi}. In Choi's paper, $\varepsilon$ was taken to be the
vacuum dielectric constant ($\varepsilon$ =1), which is not justified in our
opinion. Although there is no screening outside the atomic plane of the 2D
material, the screening nonetheless exists within the plane and cannot be
ignored. In general, $\varepsilon$ should depend on the electronic structure,
thus the bandgap of the 2D materials. As realistic 2D semiconductors are
quasi-two-dimensional (quasi-2D), the electrostatic potential is of $1/r$-type
as opposed to $\ln r$-type in an ideal 2D system. Therefore, it is appropriate
to treat a quasi-2D system as a 3D system with a very small out-of-plane
dimension. Furthermore, we have shown in the Supporting Material that $\mu$ is
a linear function of $E_{g}$, which is supported by experiments \cite{mit}.

\begin{figure}
[ph]\centering
{\includegraphics[width=0.5\textwidth]{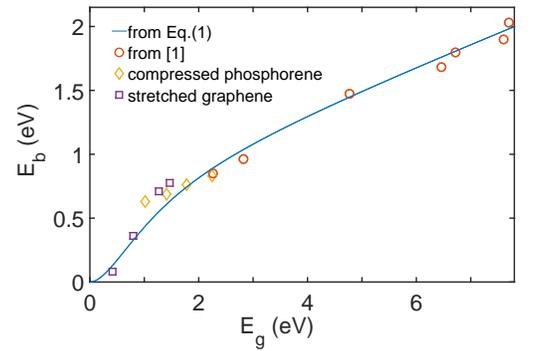}}
\caption{$E_{b}$ vs. $E_{g}$ relation determined from first-principles GW-BSE method (symbols). The circles are taken from \cite{choi}; diamonds are for compressed phosphorene and squares are for stretched graphene.
The solid curve is a fit based on Eq.(\ref{2i}).}\label{fig1}
\end{figure}

Following a simple electrostatic analysis and the harmonic oscillator model of
the static dielectric function \cite{pan}, we can express $\varepsilon$ as a
function of $E_{g}$; the frequency of the harmonic oscillator is given by
$E_{g}/\hbar$. Substituting $\mu(E_{g})$ and $\varepsilon(E_{g})$ into the
first Bohr-level of the 2D hydrogenic model \cite{yang}, we obtain $E_{b}$ for
a 2D semiconductor as following \cite{supp}:%
\begin{equation}
E_{b}=-\frac{2}{\hbar^{2}}(\frac{e^{2}}{4\pi\epsilon_{0}})^{2}\mu\left[
1+\frac{1}{2}\frac{\frac{\hbar^{2}\omega_{p}^{2}}{E_{g}^{2}}\frac{16t}{\pi
a_{0}}}{1+\sqrt{1+\frac{\hbar^{2}\omega_{p}^{2}}{E_{g}^{2}}\frac{16t}{\pi
a_{0}}}}\right]  ^{-2},\label{2i}%
\end{equation}
where $a_{0}=4\pi\epsilon_{0}\hbar^{2}/(2\mu e^{2})$. $\omega_{p}^{2}%
=n_{v}e^{2}/(\epsilon_{0}m)$, $m$ is the mass of the electron; $n_{v}$ is the
number density of the valence electrons, and $t$ denotes the thickness of the
quasi-2D semiconductor. In Fig.\ref{fig1}, we fit the analytic expression of
Eq.(\ref{2i}) to the first-principles GW-BSE results, yielding a reasonable
agreement between the two. The analytic model predicts that (i) the linear
scaling relation applies to larger bandgaps (%
$>$
2 eV); (ii) a deviation from the linear scaling relation happens for smaller
bandgaps; (iii) As $E_{g}\rightarrow0$, $E_{b}\rightarrow0$. The last
prediction is qualitatively consistent with the fact that no stable static
exciton exists in metals. Recently, an effective 2D dielectric constant has
been proposed by averaging electronic screening over the extend of the
exciton, based on which the correlation between $E_{b}$ vs. $E_{g}$ has been
examined for 51 transition metal dichalcogenides \cite{ols}. As shown in Fig.
2 of ref. \cite{ols}, the results appear to agree with our finding, i.e., a
deviation from the linear scaling is apparent for small bandgaps. We should
emphasize however that the present model is too crude to be of a predictive
power. In particular, the model does not apply to 2D semiconductors whose
bandgap is vanishingly small. In these materials, nonlocal and dynamical
screening is more important, hence the single-particle hydrogenic model is
inadequate and one has to resort to many-body approaches.\newline$^{\ast}%
$Corresponding author: ganglu@csun.edu

\end{document}


\title{Supplemental Material for \\Comment on ``Linear Scaling of the Exciton Binding Energy versus the Band Gap
of Two-Dimensional Materials''}
\author{Mingliang Zhang, Ling-Yi Huang, Xu Zhang, and Gang Lu$^{*}$}
\affiliation{Department of Physics and Astronomy, California State University Northridge,
Northridge, CA 91330}
\date{July 17, 2016}

\maketitle

\section{quasi-2D static dielectric function}

\begin{figure}
[ph]\centering {\includegraphics[width=0.5\textwidth]{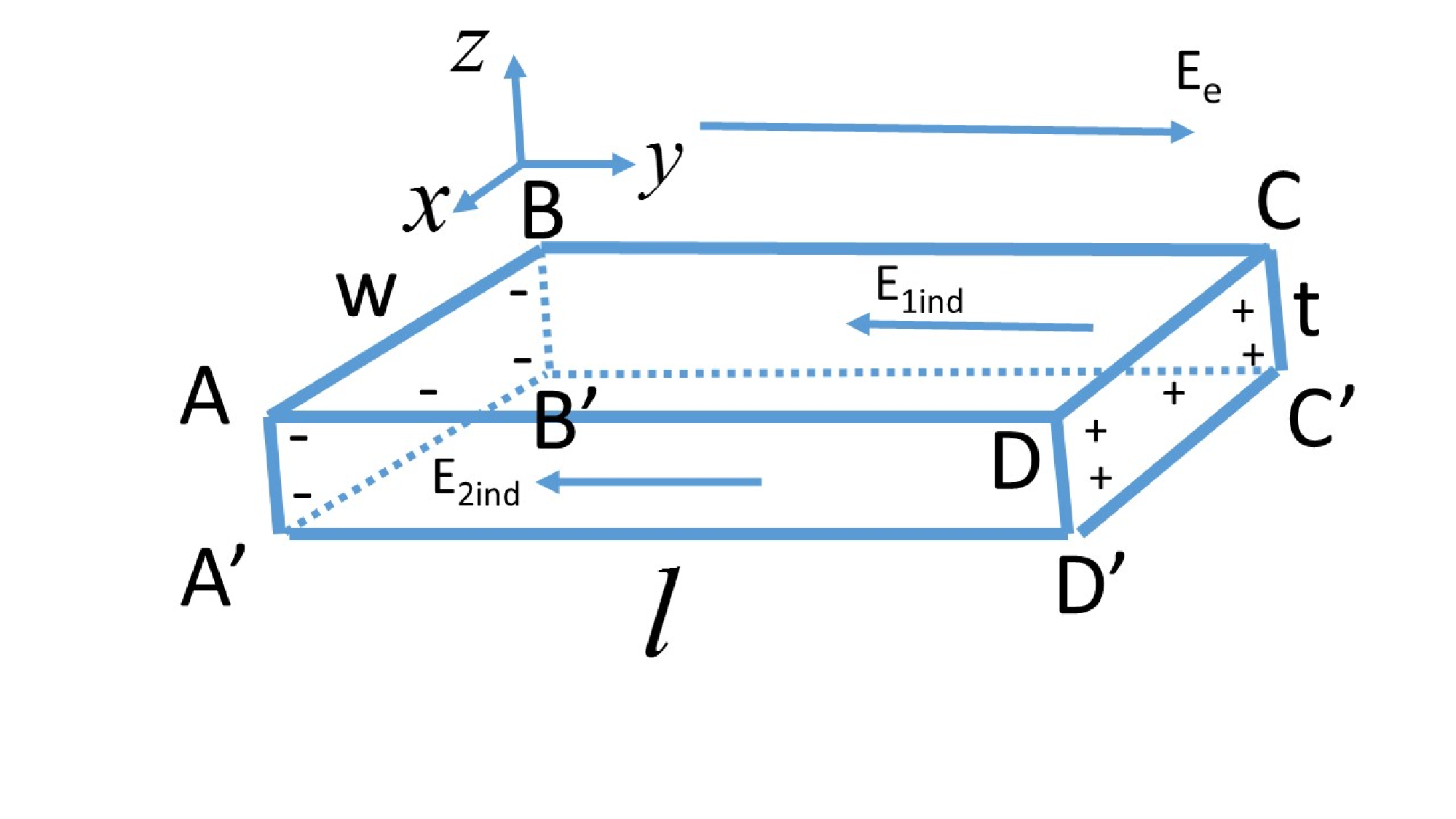}}
\caption{A quasi-2D plate with its length $l$, width $w$, thickness $t$. $E_{e}$ is the external field. $E_{1ind}$ and $E_{2ind}$ are the induced fields produced by the induced charges on the two
side surfaces $CDD'C'$ and $ABB'A'$.}\label{fig1}
\end{figure}

In a continuum approximation, a quasi-2D material can be viewed as a very thin
3D material. Thus, we may obtain $\varepsilon^{2D}$ by first considering the
dielectric screening in a 3D cuboid in Fig.\ref{fig1}, and then let the
thickness $t$ approach to zero.

Let the cuboid subject to a uniform static external field $E_{e}$ which is
along the y direction. The surface charge density $\sigma$ on $CC^{\prime
}D^{\prime}D$ is \cite{pan}%
\begin{equation}
\sigma=\chi\epsilon_{0}E, \label{sc}%
\end{equation}
where $\chi$ is susceptibility and $E$ is the total electric field at
$CC^{\prime}D^{\prime}D$. In an homogeneous external field, an ellipsoid is
uniformly polarized \cite{str}. Since a plate is approximately a degenerated
ellipsoid, the plate is uniformly polarized to a good approximation.

We next use the harmonic oscillator model for the susceptibility \cite{pan}.
Assume the valence electrons can be described by just use one type of
oscillator with eigenfrequency $\omega_{0}$. $\omega_{0}$ can be approximated
as $\omega_{0}=E_{g}/\hbar$, where $E_{g}$ is the band gap of the material.
Thus the static susceptibility is%
\begin{equation}
\chi=\frac{\hbar^{2}\omega_{p}^{2}}{E_{g}^{2}}, \label{ci}%
\end{equation}
where $\omega_{p}^{2}=n_{v}e^{2}/(\epsilon_{0}m)$; $n_{v}$ is the number
density of the valence electrons and $m$ is the mass of electron. The induced
charge $Q$ on $CC^{\prime}D^{\prime}D$ is
\begin{equation}
Q=\sigma wt. \label{ac}%
\end{equation}
There is same amount of charge with the opposite sign on $ABB^{\prime
}A^{\prime}$. For a quasi-2D system, one could consider the rectangles
$ABB^{\prime}A^{\prime}$ and $CC^{\prime}D^{\prime}D$ shrinking to two
straight-lines with length $w$. The linear charge density $\lambda$ is
\begin{equation}
\lambda=\chi\epsilon_{0}Et. \label{xd}%
\end{equation}

We next use the net local field $E$ at the center point of the plane to define
the effective dielectric function $\varepsilon^{2D}$ of the 2D material:%
\begin{equation}
\varepsilon^{2D}=\frac{E_{e}}{E}. \label{di1}%
\end{equation}
At the center, the induced field $E_{1ind}$ produced by the right line is
\begin{equation}
E_{1ind}=\frac{\chi t}{2\pi(l/2)}E. \label{1f}%
\end{equation}
A similar field $E_{2ind}$ is produced by the left line. The net field $E$ at
the center is%
\begin{equation}
E=E_{e}-E_{1ind}-E_{2ind}=E_{e}-\frac{2\chi t}{\pi l}E. \label{ne}%
\end{equation}
From Eq.(\ref{ne}), one has%
\begin{equation}
E=\frac{E_{e}}{1+\frac{2\chi t}{\pi l}}. \label{ne1}%
\end{equation}
By means of Eq.(\ref{di1}),
\begin{equation}
\varepsilon^{2D}=1+\frac{2\chi t}{\pi l}. \label{di}%
\end{equation}

A static wave-vector dependent dielectric function can be obtained from
Eq.(\ref{di}):%

\begin{equation}
\varepsilon^{2D}(\mathbf{q},0)=1+\frac{4\chi}{a^{2}q} \label{sd}%
\end{equation}
where $q=|\mathbf{q}|,$ $a$ is the characteristic length for the non-local
effect of the field. Eq.(\ref{sd}) is similar to Eq.(26) of \cite{lee}.

\section{exciton binding energy in quasi-2d system}

The exciton can be modeled by a 2D hydrogenic model:
\begin{equation}
-\frac{\hbar^{2}}{2\mu}(\frac{\partial^{2}\psi}{\partial x^{2}}+\frac
{\partial^{2}\psi}{\partial y^{2}})-\frac{1}{4\pi\epsilon_{0}\varepsilon^{2D}%
}\frac{e^{2}}{r}\psi=E\psi(x,y), \label{se}%
\end{equation}
where $r=\sqrt{x^{2}+y^{2}}$,%
\begin{equation}
\mu=\frac{m_{e}m_{h}}{m_{e}+m_{h}} \label{rm}%
\end{equation}
is the reduced mass of the electron and the hole, $m_{e}$ and $m_{h}$ are the
effective mass of the electron and the hole.

The binding energy of a ground state 2D exciton is \cite{yang}%
\begin{equation}
E_{b}=-\frac{2}{\hbar^{2}}(\frac{e^{2}}{4\pi\epsilon_{0}})^{2}\frac{\mu
}{(\varepsilon^{2D})^{2}}, \label{ev}%
\end{equation}
The first$\ $Bohr radius is
\begin{equation}
r_{1}=\frac{\hbar^{2}}{4\mu\frac{e^{2}}{4\pi\epsilon_{0}}}\varepsilon^{2D}.
\label{br}%
\end{equation}
The Bohr radius $r_{1}$ represents the electron-hole distance of the exciton,
which corresponds to $l$ of the cuboid discussed above. Combining
Eqs.(\ref{br},\ref{di}), we can determine the Bohr radius, $r_{1}$
self-consistently:
\begin{equation}
r_{1}=\frac{a_{0}}{2}(1+\frac{2\chi t}{\pi r_{1}}),\text{ } \label{r1}%
\end{equation}
where
\begin{equation}
a_{0}=\frac{\hbar^{2}}{2\mu\frac{e^{2}}{4\pi\epsilon_{0}}}. \label{a0}%
\end{equation}
The solution of Eq.(\ref{r1}) yields
\begin{equation}
r_{1}=\frac{a_{0}}{4}\{1+\sqrt{1+\frac{\hbar^{2}\omega_{p}^{2}}{E_{g}^{2}%
}\frac{16t}{\pi a_{0}}}\}. \label{r1v}%
\end{equation}
Substitute Eq.(\ref{r1v}) into Eq.(\ref{di}), we obtain the dielectric
constant for the 2D system:%
\begin{equation}
\varepsilon^{2D}(r_{1})=1+\frac{1}{2}\frac{\frac{\hbar^{2}\omega_{p}^{2}%
}{E_{g}^{2}}\frac{16t}{\pi a_{0}}}{1+\sqrt{1+\frac{\hbar^{2}\omega_{p}^{2}%
}{E_{g}^{2}}\frac{16t}{\pi a_{0}}}}. \label{2e}%
\end{equation}
Combining Eq.(\ref{2e}) into Eq.(\ref{ev}), we have the exciton binding energy
in the quasi-2D system:%
\begin{equation}
E_{b}=-\frac{2\mu}{\hbar^{2}}(\frac{e^{2}}{4\pi\epsilon_{0}})^{2}\left[
1+\frac{1}{2}\frac{\frac{\hbar^{2}\omega_{p}^{2}}{E_{g}^{2}}\frac{16t}{\pi
a_{0}}}{1+\sqrt{1+\frac{\hbar^{2}\omega_{p}^{2}}{E_{g}^{2}}\frac{16t}{\pi
a_{0}}}}\right]  ^{-2}. \label{2i}%
\end{equation}

\section{linear relation between $\mu$ and $E_{g}$}

In this section, we show that the reduced mass $\mu$ is a linear function of
$E_{g}$. Let us consider a one-dimensional system for simplicity. The
effective mass $m_{n}^{\ast}$ of either electron or hole in the $n$th band%
\begin{equation}
m_{n}^{\ast}=\frac{\hbar^{2}}{\frac{\partial^{2}E_{nk}}{\partial k^{2}}},
\label{em1}%
\end{equation}
where $E_{nk}$ is the energy dispersion relation with the wave-vector
$\mathbf{k}$ for the $n$th band. With the Tight-binding approximation, the
contribution to $E_{nk}$ is
\begin{equation}
E_{nk}=I_{n}\cos ka\text{ or }=I_{n}\sin ka, \label{dis}%
\end{equation}
where $I_{n}$ is the transfer integral of the $n$th band, and $a$ is the
lattice constant. Combining Eqs.(\ref{em1},\ref{dis}), one has%
\begin{equation}
m_{n}^{\ast}|_{\mathbf{k}=0}\thicksim\frac{\hbar^{2}}{I_{n}a^{2}}. \label{m1}%
\end{equation}

The band gap at $\mathbf{k}$ point between the $n$th band and the $(n-1)$th
band is%

\[
E_{g\mathbf{k}}^{nn-1}=E_{n\mathbf{k}}-E_{n-1\mathbf{k}}\thicksim
(E_{n}-E_{n-1})-I_{n}-I_{n-1}%
\]
where $(E_{n}-E_{n-1})$ is the energy interval between the atomic (covalent
crystal) or molecular (molecular crystal) $n$th and the $(n-1)$th energy
levels. Assume $I_{n}\thicksim I_{n-1}$, one may express the transfer integral
$I_{n}$ with respect to the band gap $E_{g}$:%
\begin{equation}
I_{n}\thicksim\frac{1}{2}[(E_{n}-E_{n-1})-E_{g\mathbf{k}}^{nn-1}]. \label{m2}%
\end{equation}
Substitute Eq.(\ref{m2}) into Eq.(\ref{m1}), we have
\[
m_{n}^{\ast}\thicksim\frac{2\hbar^{2}}{a^{2}[(E_{n}-E_{n-1})-E_{g\mathbf{k}%
}^{nn-1}]}.
\]
If
\begin{equation}
\frac{E_{g\mathbf{k}}^{nn-1}}{E_{n}-E_{n-1}}\ll1, \label{ap}%
\end{equation}
to first order of this small parameter, we have
\begin{equation}
m_{n}^{\ast}\thicksim\frac{2\hbar^{2}}{a^{2}(E_{n}-E_{n-1})}[1+\frac
{E_{g\mathbf{k}}^{nn-1}}{E_{n}-E_{n-1}}]. \label{em}%
\end{equation}
The effective mass $m^{\ast}$ is a linear function of band gap $E_{g}$. One
may notice that the approximation (\ref{ap}) is reasonable for the valence
band and conduction band, because the dispersion the band gap is smaller than
the energy interval between the two neighboring energy levels.

From Eqs.(\ref{ap},\ref{em}), the effective mass $m_{e}$ of the electron can
be written in a form:
\begin{equation}
m_{e}=b_{1}(1+\frac{k_{1}}{b_{1}}E_{g}), \label{me}%
\end{equation}
with $\frac{k_{1}}{b_{1}}<<1$. Similarly, the effective mass $m_{h}$ of the
hole can be written as
\begin{equation}
m_{h}=b_{2}(1+\frac{k_{2}}{b_{2}}E_{g}), \label{mh}%
\end{equation}
with $\frac{k_{2}}{b_{2}}<<1$. Substitute Eqs.(\ref{me},\ref{mh}) into the
definition (\ref{rm}) of the reduced mass $\mu$, one has%
\begin{equation}
\mu\thickapprox\frac{b_{1}b_{2}}{b_{1}+b_{2}}[1+(\frac{k_{1}}{b_{1}}%
+\frac{k_{2}}{b_{2}}-\frac{k_{1}+k_{2}}{b_{1}+b_{2}})E_{g}]. \label{rm1}%
\end{equation}
Eq.(\ref{rm1}) indicates that $\mu$ is a linear function of $E_{g}$.

\section{Computational details of GW-BSE}

\begin{figure}
[ph]%
\centering {\includegraphics[width=0.48\textwidth]{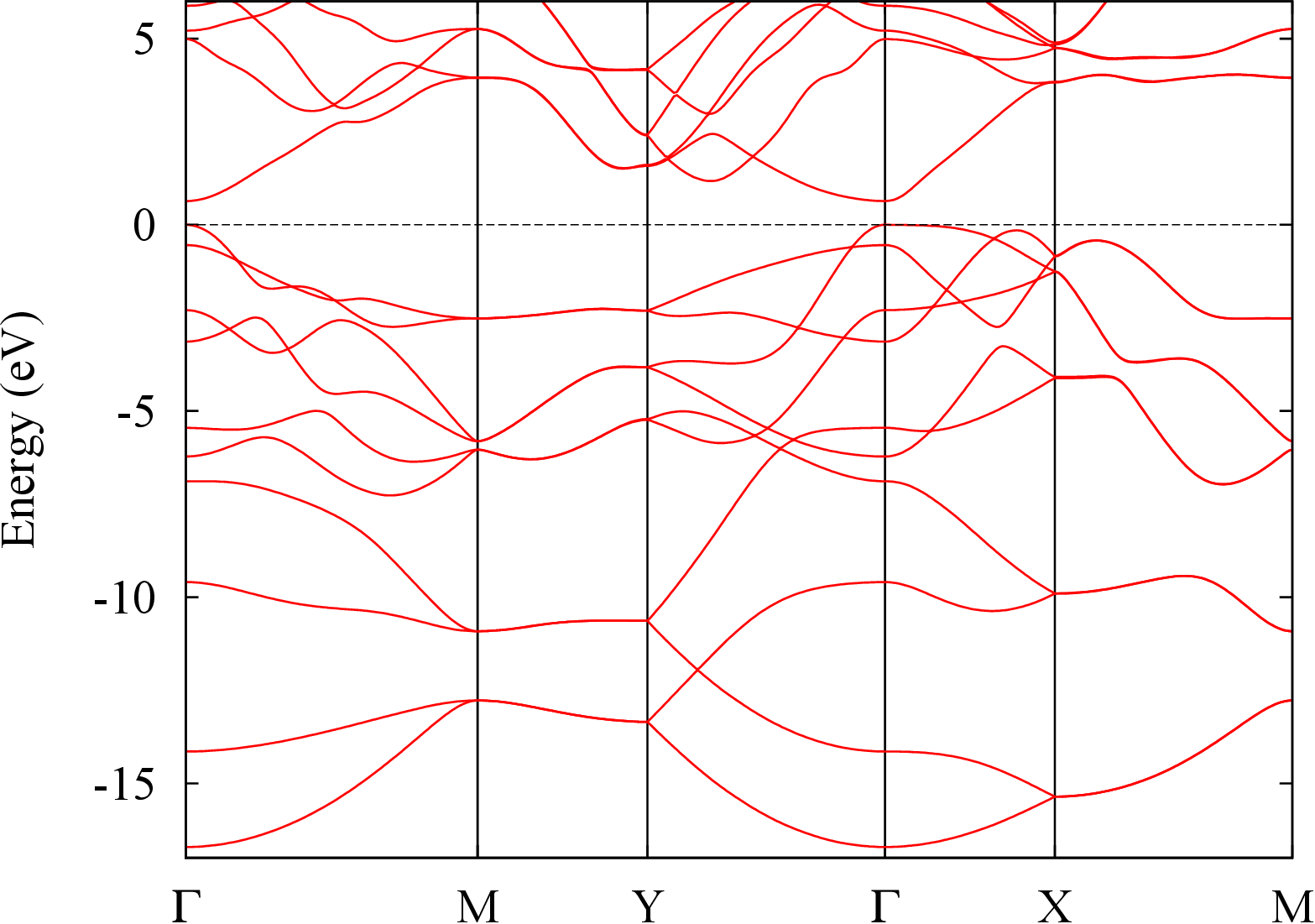}}
\caption{Calculated QP band structure of phosphorene with 93\% of relaxed lattice constants $a$ and $b$; here, we use $L_{z}=30$ \AA.}\label{bnds_gw_phosphorene_093_lz30}%

\end{figure}

The Vienna $ab$ $initio$ simulation package \cite{vasp} was used to perform
the GW+BSE calculations\cite{hyb,roh}. Our $GW$ calculations were carried out
in the partially self-consistent way, the so-called $GW_{0}$ scheme. More
precisely, we actually carried out the $G_{3}W_{0}$ calculations for both
phosphorene and graphene, where the Green's function is self-consistently
updated 3 times after the one shot $GW$. The quasi-particle (QP) band gap was
found to be converged within 0.01 eV in the $G_{3}W_{0}$ scheme. The approach
of the maximally localized Wannier functions\cite{w90} was used to plot QP
band structure. Solving the BSE on top of the preceding $GW$ results, we could
obtain the optical gap. Because we were dealing with 2D materials in which the
QP band gap depends on the spatial separation, usually denoted by $L_{z}$,
between adjacent 2D layers, we extrapolated the gap to the limit of infinite
$L_{z}$.

\begin{figure}
[ph]%
\centering {\includegraphics[width=0.48\textwidth]{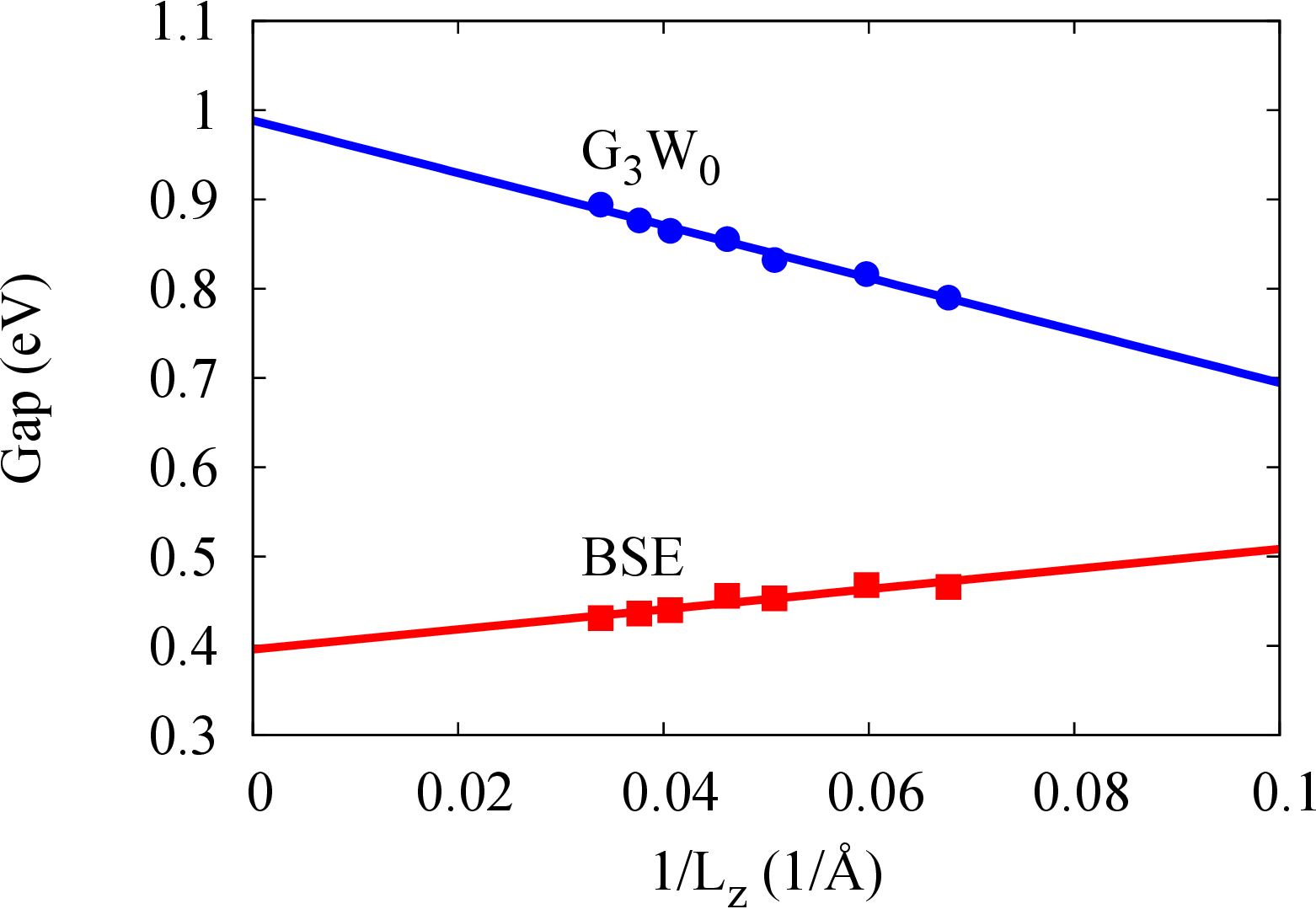}}
\caption{The QP band gap ($G_{3}W_{0}$) and optical gap (BSE)
as a function of the inverse spatial separation $1/L_{z}$ for phosphorene with 93\% of relaxed lattice constants $a$ and $b$. The dots (circles and squares) are taken  from actually calculated results while the lines represent fitting results of the corresponding dots.}\label{gap_vs_lz_phosphorene_093}%

\end{figure}

We start from fully relaxed phosphorene and then isotropically (2D
hydrostatically) compressed it up to 93\% (97\%, 95\% and 93\%) of the relaxed
lattice constants. Without compression, the optimized lattice constants are
$a=4.61$ \AA \:and $b=3.30$ \AA . We used Perdew-Burke-Ernzerfhof (PBE)
exchange-correlation functional and PAW pseudopotentials, and an energy
cut-off of 400 eV. An $11\times15\times1$ $k$-point mesh was used, and 192
unoccupied bands were sufficient to get convergent QP band gaps. We obtained
basically identical results as shown in \cite{zz} for zero compression, so we
won't redundantly show them here. As for compressed phosphorene,
it still has the direct band gap at the $\Gamma$-point but the band gap
becomes smaller. In Fig.\ref{bnds_gw_phosphorene_093_lz30}, we show the QP
band structure from the $G_{3}W_{0}$calculation with $L_{z}=30$ \AA for
phosphorene with 93\% of the relaxed lattice constants. The corresponding QP
band gap ($G_{3}W_{0}$) and optical gap (BSE) as a function of the inverse
spatial separation $1/L_{z}$ are shown in Fig.\ref{gap_vs_lz_phosphorene_093}.



\begin{figure}
[ph]%
\centering {\includegraphics[width=0.48\textwidth]{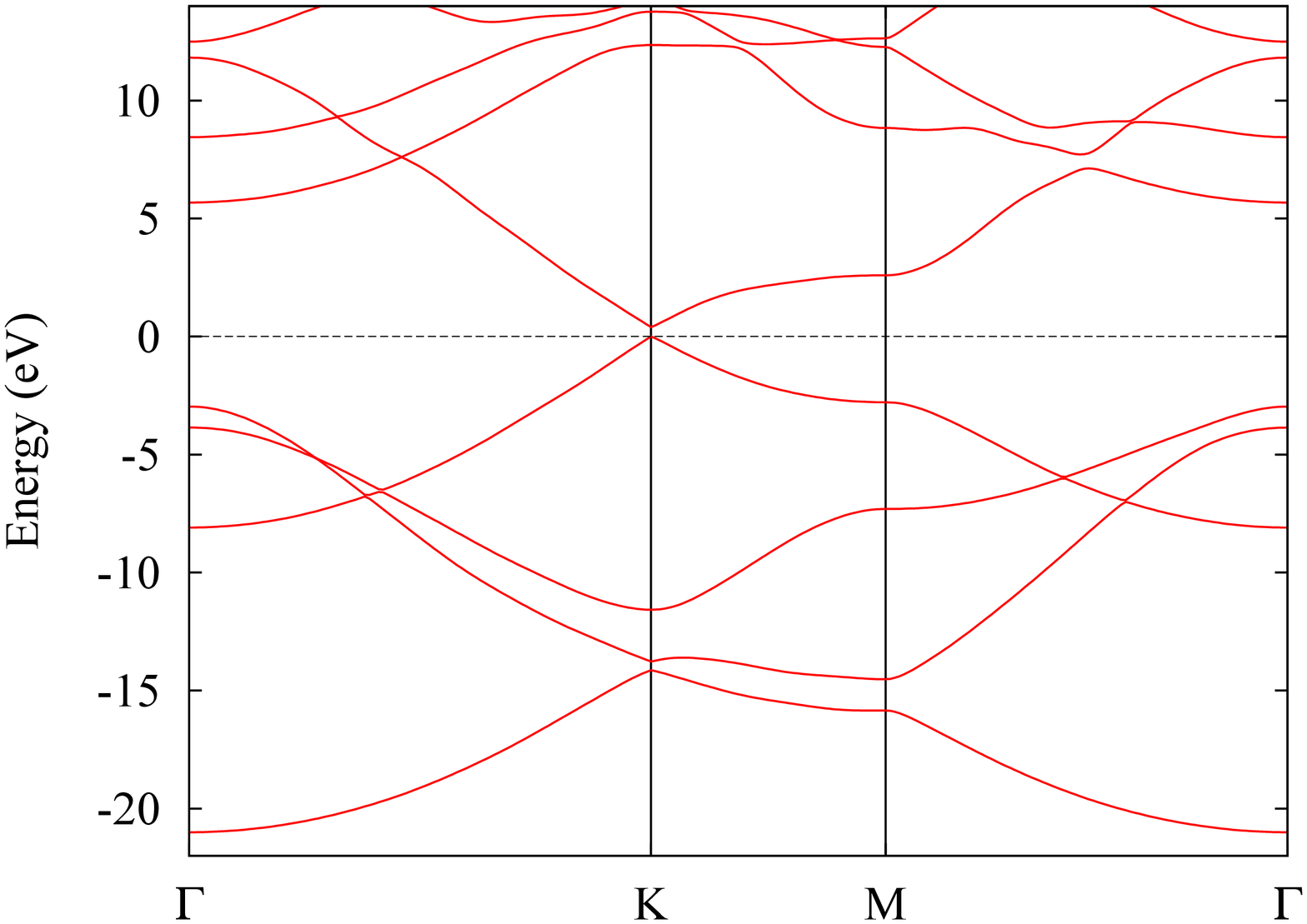}}
\caption{Calculated QP band structure of 1\% strained graphene with $L_{z}=30$ \AA.}\label{band_structure_graphene_101_lz30}%

\end{figure}

\begin{figure}
[ph]%
\centering {\includegraphics[width=0.48\textwidth]{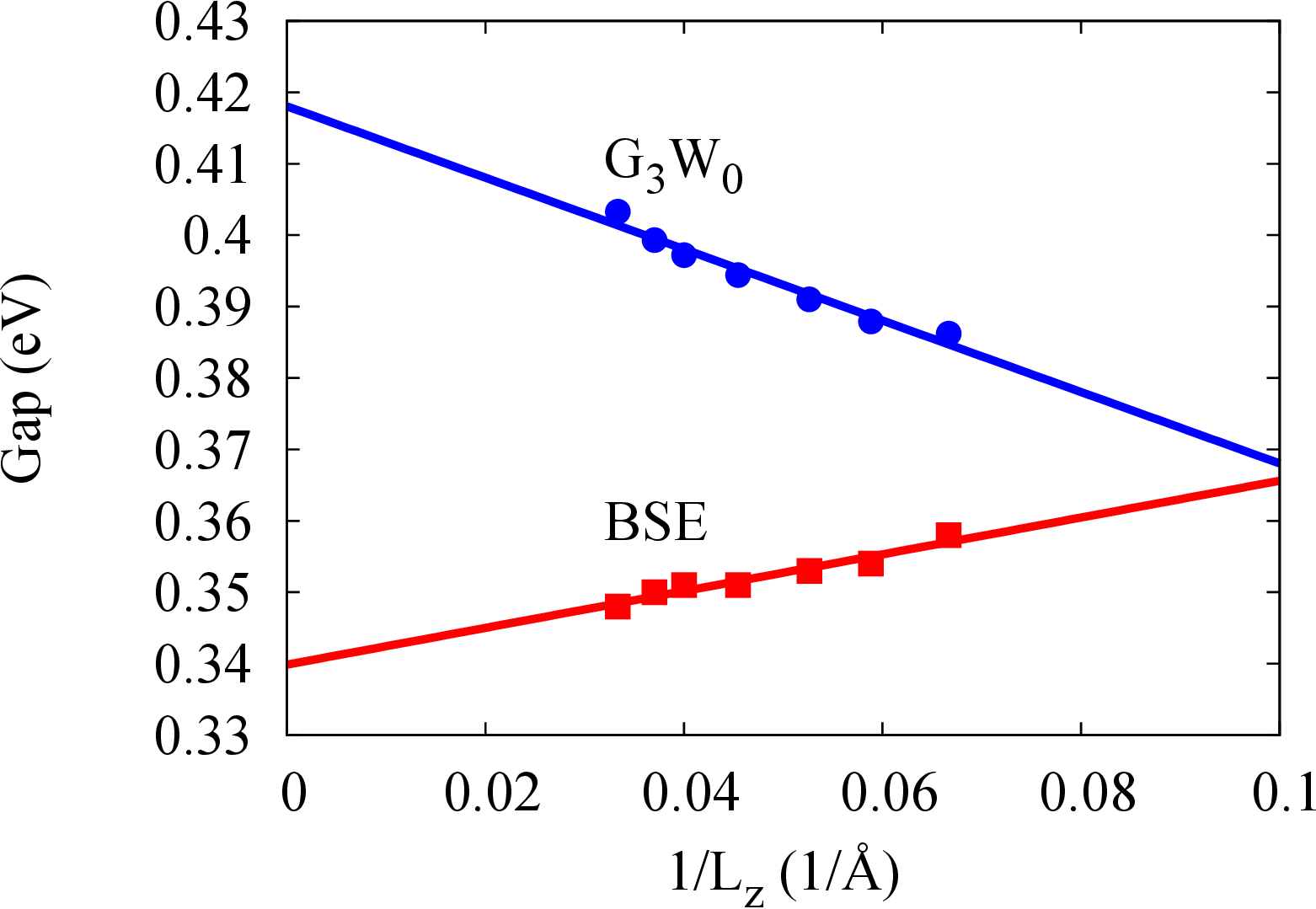}}
\caption{The QP band gap ($G_{3}W_{0}$) and optical gap (BSE)
as a function of the inverse spatial separation $1/L_{z}$ for 1\% strained graphene. The dots (circles and squares) are taken from actually calculated results while the lines are the fits to the corresponding data points.}\label{Eg_vs_Lz_graphene_101}%

\end{figure}

Next, in order to investigate the relationship between $E_{b}$ and $E_{g}$ in
the very small gap region, we tried to apply strains on graphene, which is a
zero band gap 2D material. We applied small strains (1\%, 1.5\%, 2\%, and
2.5\%) along the direction of one primitive vector and maintained the area of
a unit cell fixed. We started with experimental structure in which the C atoms
are about 1.42 \AA \:apart when zero strain is applied. We again used PBE and
PAW pseudopotentials with an energy cut-off of 408 eV. A $15\times15\times1$
$k$-point mesh and 1320 unoccupied bands were used in our calculations. For
graphene under 1\% strain, calculated QP band structure, from the $G_{3}W_{0}$
calculation with $L_{z}=30$ \AA , is shown in
Fig.\ref{band_structure_graphene_101_lz30}. The QP band gap ($G_{3}W_{0}$) and
optical gap (BSE) as a function of the inverse spatial separation $1/L_{z}$
are shown in Fig.\ref{Eg_vs_Lz_graphene_101}. \bigskip\newline*Corresponding author:~ganglu@csun.edu

